# On filamentation in the dense plasma focus

S K H Auluck[1]

International Scientific Committee for Dense Magnetized Plasmas,
http://www.icdmp.pl/isc-dmp
Hery 23, P.O. Box 49, 00-908 Warsaw, Poland

Abstract: Striking pictures showing filamentary structures in plasma focus have intrigued researchers from the early days of plasma focus research. A definitive understanding of their occurrence, origin, structure and role in plasma focus physics is still not in sight as summarized in a recent comprehensive review. This is because they are often not observed in a "standard mode" of plasma focus operation with pure deuterium, particularly in large installations, but are found in smaller experiments or those with gaseous admixtures. This has led to the suspicion that filaments are not a native feature of the plasma focus phenomenon. Recent success in observation of filaments in PF-1000 in a pure deuterium operation by a novel modification of the interferometer system that allows simultaneous interferometry and schlieren photography changes this situation. This Letter looks at the implications of this development in the larger context of plasma focus physics. Conceptualization of filamentation as a native feature of the travelling current distribution behind an ionizing strong shock wave is shown to be a feasible paradigm that can be formulated as a computable model for filamentation in the plasma focus.

Striking pictures of filamentary structures in the plasma focus [1] have intrigued researchers from the early days of plasma focus research [2]. A recent review of the scientific status of the plasma focus [3] has summarized all that is known about this phenomenon as well as theories and conjectures about what it might be. Many researchers invoke [3] conjectures partially supported by experiments to conclude that they are local current concentrations. Another view represents them as vortex filaments [1,4]. Direct observation of filaments in large, better diagnosed plasma focus facilities is mostly absent in pure deuterium shots but are reported in shots with gaseous admixtures [3]. This has led to the suspicion that they might represent some deviation from normal plasma focus operation triggered by the change in the average atomic number of the gas, perhaps via tearing modes [5].

---

[1] Corresponding Author. email: skhauluck@gmail.com

This situation has changed with the recent success in direct observation of filaments in PF-1000 using a novel diagnostic that enables simultaneous recording of an interferogram and a schlieren picture [6]. Although only preliminary results have been released so far, their significance in the overall scheme of plasma focus physics cannot be overemphasized. This paper briefly summarizes the context provided by the current state of understanding before discussing the implications of the new findings.

The present paradigm of plasma focus research can be succinctly summarized in terms of a Standard Narrative [3]. Its chief defining characteristic feature is a travelling distribution of current (that can be initiated by a variety of mechanisms [3], but is most commonly initiated by a sliding discharge on an insulator) accelerated by its own magnetic field that drives an ionizing shock wave into the neutral fill gas, mainly deuterium, and its subsequent convergence into a dense and hot plasma column. Physical separation of the current distribution (measured using magnetic probes) and the dense shockwave plasma (that is seen in most emissivity and refractivity-based diagnostics) is a well-established fact. Data available till now do not clearly indicate whether the filaments are localized within the current distribution (which is usually not visible in emissivity/refractivity-based diagnostics) or the dense sheath or at their boundary. Filaments seen in plasma focus devices are typically of sub-millimeter diameter while the bulk plasma is of few mm diameter. The diagnostic technique (whether emissivity based such as visible or soft x-ray framing or refractivity based such as interferometry, schlieren or shadowgraphy) usually shows the position of the filament along with some sort of outline of the gross plasma. These pictures support conjectures that the filaments are local concentrations of current or local Bennett equilibria formed following some variety of instability [3]. Azimuthally distributed erosion marks on the anode are also considered as indication of presence of filamentary current concentrations [7]. Observations of magnetic field component along the filament axis [8] and the quantitative agreement between the scale length of the filament and the Hall MHD scale length $c/\omega_{pi}$ [9] have been interpreted [4] as evidence of a vortex-like structure perhaps resulting from Turner relaxation [10].

All these conjectures and theories have one common theme: they present the filamentation as a phenomenon that is separate from the main feature of plasma focus phenomenon, that consists of transport of a current distribution as a structure travelling through neutral gas behind an ionizing shockwave.

The preliminary results from PF-1000 experiments previewed [6] in the Annual Meeting of the International Scientific Committee on Dense Magnetized Plasmas (published online) differ from all the previous experiments in three respects.

Firstly, they simultaneously look at the density structure of the pinch zone using two refractivity-based techniques, sensitive to the change in line-of-sight-integrated refractivity and its gradient respectively, utilizing the same laser pulse. The technique sensitive to refractivity (interferometry) looks at features of the larger plasma which misses the smaller features while the one sensitive to its gradient (schlieren) looks at smaller scale features but does not show features of the larger scale structure. Their novel combination provides a contextual significance to both the techniques that is not available to the two techniques used separately.

Secondly, they show the first ever glimpse of structure in the low-density region immediately outside the dense pinch zone that is shown (using magnetic probes and interferometry: see Fig 86, Ref 3 and associated text) to be the region where current is localized in the pinch phase. This structure unmistakably consists of filaments that are of sub-millimeter diameter (see figure b on page 7 of Ref 6). The filaments are seen to form a 3-D network that is also present at the boundary of the dense column. Significantly, the preliminary report indicates relative absence of observable filaments in the interior regions of the dense column and at the location of closed interference fringes that are interpreted as bounded plasmoid structures.

The third significant distinction of these results is that they are obtained with pure deuterium "standard mode" of plasma focus operation [3] in a mega ampere device. While filaments are observed with gas admixtures in PF-1000[11], they appear to be differently oriented with respect to the sheath structure than the classic [1] filaments in plasma focus.

Descriptively, the filamentary structures seen in the schlieren pictures in the boundary and outer regions of the dense plasma appear to span their location densely while those in the interior regions are relatively well-separated, as would be the case when they are mainly located in an annular region whose 2-D projection would have them appear close together in the peripheral regions but not in the interior regions. They appear as elongated structures which bend smoothly. There are apparent "end points" and "junctions" which could be an effect of projecting 3-dimensional filamentary structures onto a 2-dimensional image.

The significance of this observation is that this kind of dense filamentary network structure cannot be described as an acquired property of a non-filamentary, smooth current distribution driving an ionizing shockwave into neutral gas by the operation of a subsequent phenomenon such as an instability or turbulence. The idea of instability assumes a pre-existing configuration with the unstable mode absent. Then, one (or a small number) of eigenmodes grow on account of a feedback mechanism that is built into the dynamical equations. The feedback mechanism many times involves some sort of phase matching condition that allows a steady flow of excess free energy into the motion represented by the unstable mode. Such phase matching conditions involve the mode numbers that identify the unstable mode and can therefore be satisfied only under special conditions by a small number of modes. The well-observed phenomenon of m=0 instability is a good illustration. It is preceded by a smooth, often symmetric, plasma structure. On some devices and certain operating regimes [12], the m=0 instability is "a rare occurrence" [13]. Sometimes, it is accompanied by an m=1 helical deformation as well [3].

Simultaneous growth of a large number of modes is usually encountered in the case of fluid turbulence. But these modes have a random character and although they form coherent structures variously referred as eddies, vortex cells, turbulence packets etc., some kind of threshold condition is usually involved with their occurrence.

Generation of structure is many times looked upon as an example of self-organization [14] in open systems, where instability at critical values of external control inputs leads to growth of a few "order parameter" modes which impose their imprint on to the system via an "enslavement" mechanism.

The dense network of filaments observed recently in PF-1000 appears to be well-embedded in a larger structure and in fact seems to be aligned with its gross features. Such dense network of filaments aligned with a larger structure needs to be conceptualized as an integral, native feature of the plasma focus phenomenon, rather than as an add-on feature that may or may not occur in certain devices or operating regimes that fail to fulfill some phase matching condition or crossing of a threshold condition. It does not conform with the idea of instability described above, where only a few modes grow or of turbulence where a large number of modes grow randomly after crossing a threshold.

This is contrary to the spirit of all the theories and conjectures regarding filamentation in the plasma focus [3] which hold filamentation and plasma focus as separate phenomena. At present, there is not enough experimental information on the nature of the filamentation phenomenon in the plasma focus to construct a theory of self-organization [14].

Nevertheless, this shift in paradigm (of considering filamentation as an integral, native feature of the plasma focus phenomenon) can be put on a first-principles foundation. It can be shown [15] that Chandrasekhar-Kendall functions $\vec{\chi}_{m\kappa\gamma s}(r,\theta,z)$

$$\vec{\chi}_{m\kappa\gamma s}(r,\theta,z) = k^{-2} \exp(im\theta - i\kappa z) \begin{pmatrix} i\frac{1}{2}\gamma \hat{r}\left((sk+\kappa)J_{m+1}(\gamma r) + (sk-\kappa)J_{m-1}(\gamma r)\right) \\ +\frac{1}{2}\gamma \hat{\theta}\left((sk+\kappa)J_{m+1}(\gamma r) - (sk-\kappa)J_{m-1}(\gamma r)\right) \\ +\hat{z}\gamma^2 J_m(\gamma r) \end{pmatrix} \quad (1),$$

their generating function $\psi_{m\kappa\gamma}(r,\theta,z)$ (which is also the eigenfunction of the Laplacian)

$$\psi_{m\kappa\gamma}(r,\theta,z) = J_m(\gamma r)\exp(im\theta - i\kappa z) \quad (2)$$

and its gradient $\vec{\nabla}\psi_{m\kappa\gamma}$ form a complete orthonormal basis set of functions over infinite domain containing the axis for solenoidal, scalar and irrotational fields respectively. This enables the transformation of dynamical equations of any continuum model of the plasma (ideal MHD, resistive MHD, Hall MHD, Two-Fluid, Vlasov, Boltzmann, "fully kinetic", etc.) into cylindrical Fourier (or mode-number) space over infinite domain. The finite size of the plasma provides a physical lower limit to the mode number spectrum. The physical phenomena included in the model are manifested in terms of characteristic scale lengths where microscopic phenomena provide a physical upper limit on mode numbers.

Non-linear terms of the dynamical equations in mode number space provide cross-coupling between modes [15]. Modes with mode numbers $(m',\kappa',\gamma')$ and $(m'',\kappa'',\gamma'')$ cause changes in the rate of change of spectral density of the field (which may be a scalar field like density, temperature or pressure, a solenoidal field like magnetic field or a combination of solenoidal and irrotational fields like velocity, current and electric field.) at mode numbers $(m,\kappa,\gamma)$ where $m = m' \pm m''$, $\kappa = \kappa' \pm \kappa''$, $\gamma = \gamma' \pm \gamma''$. As a result, the initial spectral density distribution in mode number space of every field would tend to cascade towards smaller and larger mode numbers until the physical lower and upper limits are approached and would then become quasi-stationary. These limiting eigenmodes would modulate the gross field, such as density or current density with their characteristic spatial structure at the upper and lower limits of mode numbers.

This is a general phenomenon, independent of the details of the plasma model that is being considered and the kind of device that is being considered. It differs from the idea of instability in the sense that all possible modes interactively play a role rather than only a few. It differs from turbulence in the sense that there is no obvious threshold crossing involved and randomness of individual interacting modes gives rise to non-random, coherent outcomes [16] via nonlinear terms with phase correlation between interacting terms enforced via dynamics.

In the plasma focus case, the relevant scale lengths are related to the pinch radius and height on the gross scale and the collisionless skin depth $c/\omega_{pi}$ on the finer scale. A popular estimate of pinch radius and height normalized to anode radius a is obtained by Lee and Serban [17] from a scaling model as $\tilde{r}_p \equiv r_p/a \sim 0.12$, $\tilde{z}_p \equiv z_p/a \sim 0.8$. Actual interferograms show the radius and height of the pinch normalized to anode radius to be $\tilde{r}_p \approx 0.07$ and $\tilde{z}_p \approx 0.26$ for PF-1000 [18]; $\tilde{r}_p \approx 0.12$ and $\tilde{z}_p \approx 0.43$ for POSEIDON [19].

From (2), the lower limit $\gamma_L$ is given by $J_0(\gamma_L r_p) = 0$ so that $\gamma_L = j_{0,1}/r_p$. Similarly, the lower limit $\kappa_L$ is given by $\kappa_L = \pi/z_p$. The lower limit of azimuthal mode number is obviously m=0.

Before considering the upper limits of mode numbers, it is necessary to appreciate the nature of the eigenfunction (2), written in the real form

$$\psi_{m\kappa\gamma}(r,\theta,z) = J_m(\gamma r)\cos(m\theta - \kappa z) \qquad (3)$$

in the midplane $z = 0$. It is not only the eigenfunction for scalar fields but also represents the z-component of the eigenfunction (1) of solenoidal fields: in particular the axial magnetic field, the axial electric field, the axial current and axial vorticity.

A contour map of (3) for various values of m is shown in Fig 1. It shows islands of positive values interspersed with islands of negative values laid out in concentric rings. The radius $R_m$ of the innermost ring passing through the centers of these islands depends [20] on m as

$$\gamma R_m = 1.88834 + 1.01808\ m \qquad (4)$$

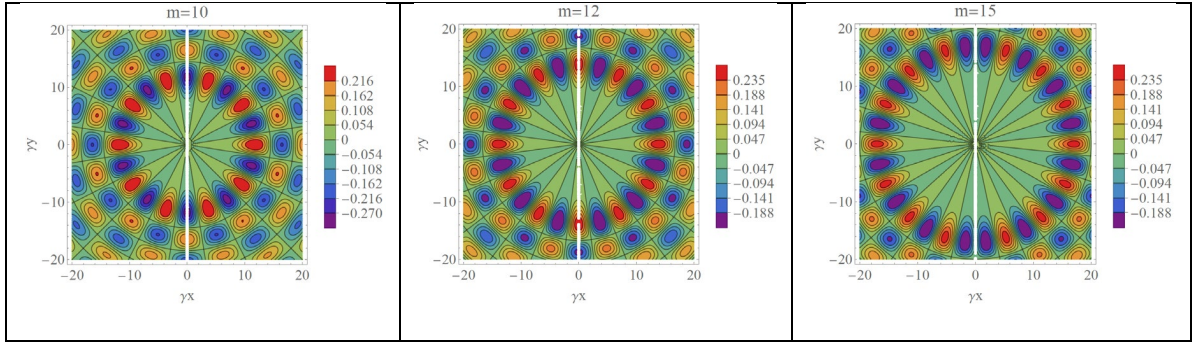

Fig 1: Contour map of the function (3), which represents the eigenfunction of scalar fields such as density, temperature and pressure and the z-component of solenoidal fields: axial magnetic field, axial current density, axial electric field and axial vorticity. Islands of positive value (yellow to red) are positioned on concentric rings interleaved with islands of negative value (green to blue).

There is sufficient similarity between the contour maps in Fig 1 and the iconic photographs of filaments published by Bostick and colleagues [1] to propose the conjecture that the higher m-value C-K functions are a viable model for filamentation. The negative-valued islands in Fig 1 would imply cavities in the density, or azimuthal "corrugations" in the gross features of the plasma sheath [1]. A cavity in density could also arise from a vortex flow. The positive valued islands of Fig 1 would tend to have local maxima in density, indicating plasma being gathered by a local pinching action of current density flowing in these structures. Such vortex flows and current concentrations would be local zones of intense energy dissipation, either through viscosity or through resistivity. As a result, the number of visible structures would be twice the azimuthal mode number.

In this paradigm, the idea that the filaments represent local current concentrations is supported because extrema of axial current density also coincide with extrema of the density. The conjecture that the filaments are vortex filaments would also be supported in this framework because the z-component of vorticity is again of the form given by (3) and shown in the figure. The claims of a component of magnetic field along the filament axis, which are based on meager experimental results [3], also seem to be supported in this paradigm in the sense that the density extrema coincide with extrema of axial component of the magnetic field.

The density profile of a single filament has been determined [4] by Abel inversion of fringe deflection in an interferogram for a small plasma focus. Its radius at half-maximum is found to be closely matching the value $c/\omega_{pi}$ determined from its peak density [4]. This suggests that the upper mode number limits, $\gamma_U, m_U, \kappa_U$ are proportional to the reciprocal of the scale length $c/\omega_{pi}$ which depends on the local ion density $n_i$ of the current carrying annular region just outside the dense plasma region where the filaments are located: $\gamma_U = f_\gamma \omega_{pi}/c$, $\kappa_U = f_\kappa \omega_{pi}/c$, $\omega_{pi}^2 = n_i e^2/\varepsilon_0 m_i$. The constants $f_\gamma, f_\kappa$ can be taken to be the difference between the first peak and first zero of functions $J_m$ and cosine respectively. It is found [20] that $f_\kappa = \pi/2$ and

$$f_\gamma \approx 2.35799 + 0.0400835m - 0.00012258m^2, \tag{5}$$

Following the qualitative understanding of PF-1000 results concerning location of the filaments in an annular region surrounding the dense plasma column, the radius of the inner circle in Fig 1, parametrically quantified by (4), can be taken to represent the boundary of the dense pinch so that the value of $m_U$ can be deduced from

$$\gamma_U r_p = 1.88834 + 1.01808\, m_U = \left(2.35799 + 0.0400835 m_U - 0.00012258 m_U^2\right) \tilde{r}_p \left(\omega_{pi} a/c\right) \tag{6}$$

This may be used to estimate the upper limit of the azimuthal mode number as the integer nearest to the solution of equation (6).

There is no good measurement of $n_i$ as yet. But a reasonable estimate can be made from the observation that 1 cm of fringe separation implies a density $\sim (1-3) \times 10^{23}\,\text{m}^{-3}$ [21]. The region immediately outside the dense column has fringes separated by 2-5 mm. A fair estimate for this number might be $n_i = P \times 10^{24}\,\text{m}^{-3}, P \sim 1$.

For PF-1000, a=0.115 m, $\tilde{r}_p \sim 0.07$, $\tilde{r}_p (\omega_{pi} a/c) \approx 25\sqrt{P}$. Estimates of $m_U$ from (6) for various values of P (ion density in units of $10^{24}\text{m}^{-3}$) between 0.5 and 1.5 and corresponding values of $f_\gamma$ from (5) indicate that $m_U$ lies in the range 80 to 150 while $f_\gamma$ varies from 4 to 6.

The result of this exercise is a computable model of the filamentary structure. Taking as an example $m_U \simeq 100$ (for $n_i = 0.65 \times 10^{24}\,\text{m}^{-3}$), $f_\gamma = 5.14$, $\gamma_U r = 20(\tilde{r}/\tilde{r}_p)$, $\kappa_U z = 31.66\tilde{z}/\tilde{r}_p$, a 3D contour plot of eigenfunction (3) can be obtained but is unfortunately not a graphical object that yields many features to visual observation. The function in question is

$$\psi^U_{mκγ}(r,\theta,z) = J_{100}\left(20\tilde{r}/\tilde{r}_p\right)\cos\left(100\theta - 31.66\tilde{z}/\tilde{r}_p\right) \tag{7}$$

where the r and z coordinates are scaled to the radius of the plasma column. The following features can, however, be easily inferred:

1. There are m=100 pairs of filaments each pair corresponding to the values ±1 of the cosine term (supporting claims of Bostick [8] regarding "pair production" of filaments). This accords well with the recent schlieren photographs of a dense network of filaments.
2. Insofar as the eigenfunction (3) is identical with the z-component of the C-K function (1) which should represent axial magnetic field, axial component of vorticity and axial component of current density, the conjectures concerning the observed structures being current filaments [3] and vortex filaments [4,8] are both seen to be simultaneously valid. The neighboring filaments then have oppositely directed current, axial magnetic field and vorticity. Note that they are only modulations of the respective fields, not their total values. For example, although these fields have opposite signs for the neighboring filaments, the absolute magnitude of the field may not have a null point in the intermediate space.

3. The variation with z is periodic, explicitly giving rise to a helical form to the filaments that winds around the axis. Seen in a projection perpendicular to the axis, the filaments should appear at an angle to the axis: a feature clearly seen in the PF-1000 schlieren pictures.

4. Filaments in PF-1000 would have a radius of the order of $c/\omega_{pi} \sim 0.3 \text{ mm}/\sqrt{P}$ where P is the local density in units of $10^{24} \text{ m}^{-3}$.

One could test this conjecture by comparing its predictions with Bostick's work [22]. His plasma focus had a 3.4 cm diameter anode. From space and time resolved spectroscopic measurements of Hβ profile, a density $\sim 10^{24} \text{m}^{-3}$ has been reported. This makes the factor $\tilde{r}_p \left( \omega_{pi} a / c \right) \approx 6$. Equation (6) gives $m_U \sim 9$. Fig 3 of Ref 22 shows 18 filaments.

In conclusion, the recent experimental confirmation that densely packed sub-millimeter size filaments do exist in a large mega-ampere scale plasma focus operating with pure deuterium and are mainly located in the current carrying layer just outside the dense plasma column suggests that filaments need to be treated as a native feature of the plasma focus phenomenon that involves transport of a current distribution as a structure travelling through neutral gas behind an ionizing shockwave. A computable model of the filamentary structure results from the proposed conjecture that higher azimuthal mode number eigenfunctions of the Laplacian represent the filamentary structure following an argument based on condensation of spectral density of continuum model fields at physically allowed extremes of mode numbers [15].

The author would like to acknowledge Prof. Pavel Kubes for providing a copy of his results and presentation.

Data Availability Statement: "Data sharing not applicable – no new data generated"